\documentclass[aps,groupedaddress]{revtex4}
\usepackage{epsfig}
\usepackage{graphicx}
\usepackage[T1]{fontenc}
\usepackage{ae}
\usepackage{color}
\usepackage[latin1]{inputenc}
\usepackage{amssymb,amsbsy,amsmath}
\usepackage{bbm}

\begin{document}



\title[Floquet theory of the two-level system]
      {Floquet theory of the analytical solution of a periodically driven
      two-level system}
\author{Heinz-J\"urgen Schmidt$^1$, J\"urgen Schnack$^2$, and Martin Holthaus$^3$
}
\address{$^1$  Universit\"at Osnabr\"uck,
Fachbereich Physik,
 D - 49069 Osnabr\"uck, Germany\\
$^2$  Universit\"at Bielefeld,
   Fakult\"at f\"ur Physik,   D-33501 Bielefeld, Germany\\
$3$ Carl von Ossietzky Universit\"at,
Institut f\"ur Physik,
D - 26111 Oldenburg, Germany }


\begin{abstract}
We investigate the analytical solution of a two-level system subject to a monochromatical, linearly polarized external field
that was published a couple of years ago.  In particular, we derive an explicit expression for the quasienergy.
Moreover, we calculate the time evolution of a typical two-level system over a full
period by evaluating series solutions of the confluent Heun equation. This is possible
without invoking the connection problem of this equation since the
complete time evolution of the system under consideration can be reduced to that of the first quarter-period.
\end{abstract}

\maketitle
\section{Introduction}\label{sec:Intro}

There is hardly any nutshell-like model in theoretical physics which  has
been so successful in both explaining experimental observations and providing
ground-breaking conceptual insight as the two-level system exposed to a
linearly polarized time-periodic driving force. In the context of magnetic
resonance, this system has led to the development of the famous rotating wave
approximation~\cite{RabiEtAl54}, and to the unambiguous identification of
those effects which are not covered by this approximation, such as the
Bloch-Siegert shift~\cite{BlochSiegert40}. The theoretical discussion and
experimental observation by Autler and Townes of the Stark effect exhibited by
an effective two-level system in rapidly varying fields~\cite{AutlerTownes55}
by now has matured into the understanding of the ac Stark shift of atomic
and molecular energy levels in intense laser fields. Applied to the problem
of scattering of light by atoms, the driven two-level system underlies the
notion of the Mollow triplet~\cite{Mollow69,GroveEtAl77}. Shirley's profound
discussion of the solutions to the time-dependent Schr\"odinger equation of
a linearly driven two-level system~\cite{Shirley65}, based on a systematic
use of the Floquet theory for differential equations with periodic
coefficients~\cite{Floquet83,YakubovichStarzhinskii75}, already contains many
elements which were encountered again later when setting up the general
framework of quasienergies and Floquet states for periodically time-dependent
quantum systems~\cite{Zeldovich66,Ritus66,Sambe73,FainshteinEtAl78,HillEtAl16}. Indeed,
a comprehensive exposition of the mathematics of the periodically driven
two-level system, and of its ramifications for laboratory physics, easily
fills a textbook~\cite{AllenEberly75}.

The recent discovery of closed analytical solutions for monochromatically
driven two-level systems~\cite{ML07,XieHai10} therefore deserves particular
attention. Unfortunately, these formal solutions are expressed in terms of
 confluent Heun functions, which to physicists are far less familiar than
the common hypergeometric functions or confluent hypergeometric functions,
say. Thus, it still remains necessary to explore in detail how the new Heun
solutions lend themselves to a deeper understanding of the actual physics,
how known approximations can be recovered, and whether still unknown relations
can be found. The present work is intended as a first step in this direction.

The mentioned analytical solutions also hold for the cases where
the angle between the constant field and the linearly polarized one is arbitrary~\cite{ML07,XieHai10}.
These solutions bear on a transformation of the Schr\"odinger equation into a special confluent Heun differential equation.
This differential equation resembles the Razavy equation (A2) in \cite{R80} that was derived in the
context of soluble one-dimensional Schr\"odinger equations with a bistable potential.
A similar approach has  been applied to the two-level system subject to a magnetic pulse \cite{JR10,JR10a,IG14,ISI15} and to the quantum Rabi problem
\cite{ZXBL13,MPS14,XZBL17}. In these papers also the general Heun equation has been employed, see \cite{X18} for a recent survey.
In the present context it is interesting that the latter reference \cite{X18} shows how to reduce the Schr\"odinger equation of the Rabi problem
with elliptic polarization to a general Heun equation (case III in \cite {X18}).

In this paper we will reconsider the analytical solution of the linearly polarized Rabi problem
in detail and address questions connected with the Floquet theory of this problem,
namely the complete time evolution and the quasienergy of the two-level system under consideration.
Note that for this problem we have three physical parameters, the frequency $\omega$ of the monochromatical driving, the Larmor frequency $\omega_0$ of the constant magnetic field and the amplitude $F$ of the linearly polarized field.
The transformation $t\mapsto z=\sin^2 \frac{\omega t}{2}$, see (\ref{T2}), maps half the time period $[0,T/2]$ onto the range $[0,1]$ of the argument $z$
of the confluent Heun function. This raises the question how to describe the time evolution for the remaining part $[T/2,T]$.
We solve this question in section \ref{sec:F} where we will show that it is even possible to reduce the time evolution to the first quarter-period.
In particular, the full monodromy matrix can be reduced to the ``half-period monodromy" and further to the ``quarter-period monodromy matrix."

The solutions of the  confluent Heun differential equation admit power series representations
at the singular points $z=0$ and $z=1$, resp., with a convergence radius of $R=1$.
It has been argued \cite{XieHai10,X18} that the calculation of the time evolution using confluent Heun functions would require both
power series solutions and hence a procedure to connect these two ones. As remarked in \cite{X18},
this is exactly what mathematicians call the  connection problem for solutions of second order equations.
In contrast, we have found that it is not necessary to resort to the connection problem in order to calculate the time evolution and the quasienergy.
The relevant auxiliary quantities $r$ and $\alpha$ can already be determined by the quarter-period monodromy matrix that can in turn be expressed in terms
of confluent Heun functions at the value $z=1/2$.
We thus obtain an explicit analytical expression for the quasienergy of the driven two-level system in terms of two Heun functions, see
section \ref{sec:C}. This expression can be evaluated and shown to satisfactorily approximate the numerically determined quasienergy in the domain
of $\omega>3/128$. For smaller values of $\omega$ certain parameters for the involved confluent Heun functions become too large and these functions cannot longer be accurately evaluated by truncations of their series expansions although we are still in the domain of convergence.
It is, however, possible to devise approximate solutions of the Schr\"odinger equation and corresponding expressions of the quasienergy
that hold in the adiabatic limit $\omega\rightarrow 0$, see, e.~g.~, \cite{S18}, but this topic will not be further treated in the present paper.
The other limit $\omega_0\rightarrow 0$ leads to a well-known approximation of the quasienergy valid for a relatively small constant field component
that has been used in various applications, see  section \ref{sec:L}.
We will re-derive this approximation directly from the analytical expression for the quasienergy and the corresponding limit solutions of the confluent Heun equation.
Further, the time evolution of the two-level system is analytically calculated for an example in section \ref{sec:TT} and shown to agree with the numerical result.
We close with  summary and outlook in section \ref{sec:SO}.

\section{Floquet theory and time evolution of the RPL}\label{sec:F}

The Rabi problem with linear polarization (RPL) is defined by the Hamiltonian
\begin{equation}\label{F1}
  H(\tau)=\frac{1}{2}\left(
  \begin{array}{cc}
    f\,\sin\tau & \nu \\
    \nu &- f\,\sin\tau
  \end{array}
  \right)
  \;.
\end{equation}
Here $\tau=\omega\,t$ denotes the dimensionless time, $\omega$ being the frequency of the driving field into $z$-direction, $F=f\,\omega$ its amplitude
and $\omega_0=\nu\,\omega$ the Larmor frequency of the constant magnetic field into $x$-direction.
The dimensionless period is always $T\,\omega=2\pi$.
The chosen form of $H(\tau)$ follows \cite{XieHai10} and
turns out to be most convenient for the following calculations. According to Floquet theory, the general solution of the corresponding
Schr\"odinger equation ($\hbar=1$)
\begin{equation}\label{F2}
{\sf i}\, \frac{d}{d\tau}\,\psi(\tau)=H(\tau)\,\psi(\tau)
\end{equation}
can be written as
\begin{equation}\label{F3}
 \psi(\tau)=\sum_{n=1}^{2}a_n\,u_n(\tau)\,e^{-{\sf i}\,\epsilon_n\,\tau}
 \;,
\end{equation}
with time-independent coefficients $a_n$,  Floquet solutions $u_n(\tau)\,e^{-{\sf i}\,\epsilon_n\,\tau}$,
$u_n(\tau)$ being $2\pi$-periodic,
and the dimensionless quasienergies $\epsilon_n$ satisfying $\epsilon_1+\epsilon_2=0$, see, e.~g., \cite{S18} or \cite{H16}.
Sometimes it will be necessary to also consider the quasienergy in full physical dimensions that will be denoted by
\begin{equation}\label{F3a}
 {\mathcal E}\equiv \hbar\,\omega\, \epsilon
 \;.
\end{equation}

In general, the unitary evolution matrix $U(\tau,\tau_0)$ is defined as the solution of
\begin{equation}\label{F4}
{\sf i}\, \frac{d}{d\tau}\,U(\tau,\tau_0)=H(\tau)\,U(\tau,\tau_0)
\end{equation}
satisfying the initial condition
\begin{equation}\label{F5}
 U(\tau_0,\tau_0)={\mathbbm 1}
  \;.
\end{equation}
For the two-level system the first column of $U(\tau,\tau_0)$ can be viewed as a solution $\psi(\tau)={\psi_1(\tau) \choose \psi_2(\tau)}$
of (\ref{F2}) satisfying the initial condition $\psi(\tau_0)={1 \choose 0}$. The second column of $U(\tau,\tau_0)$ then is necessarily another
solution  $\widetilde{\psi}(\tau)$ of (\ref{F2}) orthogonal to $\psi(\tau)$ and satisfying the initial condition
$\widetilde{\psi}(\tau_0)={0 \choose 1}$. This uniquely determines the form of the evolution matrix to
\begin{equation}\label{F6}
 U(\tau,\tau_0)=\left(
  \begin{array}{cc}
   \psi_1(\tau)& -\overline{\psi_2(\tau)} \\
    \psi_2(\tau)&\overline{\psi_1(\tau)}
  \end{array}
  \right)
  \;,
\end{equation}
without using the special form of the Hamiltonian (\ref{F1}). The overline indicates complex conjugation. It follows that the eigenvalues of the
``monodromy matrix " $U(\tau_0+2\pi,\tau_0)$ are of the form $e^{-2\,\pi\,{\sf i}\,\epsilon_n}$ and hence
the quasienergies can be obtained by diagonalizing the monodromy matrix, taking into account that the $\epsilon_n$ are only defined up to additive integers.

Now we take into account the special form (\ref{F1}) of the Hamiltonian and set $\tau_0=0$ such that the monodromy matrix
is written as $U(2\pi,0)$. The graph of the $\sin$-function that appears in the Hamiltonian (\ref{F1})
admits an infinite symmetry group ${\mathcal G}$ that is generated by the symmetries
\begin{equation}\label{F6a}
 \sin(\pi+\tau)=-\sin(\tau)
 \;,
\end{equation}
and
\begin{equation}\label{F6b}
 \sin(-\tau )=-\sin(\tau)
 \;.
\end{equation}
For example, (\ref{F6a}) and (\ref{F6b}) imply
\begin{equation}\label{F6c}
 \sin(\pi-\tau )=-\sin(-\tau)=\sin(\tau)
 \;.
\end{equation}

Obviously, ${\mathcal G}$ operates in a natural manner on the set of solutions of the Schr\"odinger equation (\ref{F2})
by means of (anti-)unitary operators.
First, we note that according to the transformation (\ref{F6a}) the Schr\"odinger equation (\ref{F2}) is invariant under the combined operation
\begin{equation}\label{F7}
\widetilde{\mathcal T}:
\left\{\begin{array}{c}
        {\mathcal T} =\left(\begin{array}{cc}
                              0 & 1 \\
                              1 & 0
                            \end{array} \right)\\
         \tau\mapsto \pi+\tau
       \end{array}
\right.
\;,
\end{equation}
since after the time translation $\tau\mapsto \pi+\tau$ the system feels the same magnetic field up to the sign change $f\mapsto -f$
that is, however, exactly compensated by the transposition ${\mathcal T}$.
This invariance implies
\begin{equation}\label{F8}
U(\pi+\tau,\pi)={\mathcal T}\,U(\tau,0)\,{\mathcal T}
\;,
\end{equation}
and hence the time evolution in the second half-period is completely determined by the time evolution in the first one.
This is relevant for the following sections since the transformation to the confluent Heun equation only holds for the first half-period.
In particular,
\begin{equation}\label{F8a}
U(2\pi,\pi)={\mathcal T}\,U(\pi,0)\,{\mathcal T}
\;.
\end{equation}
Further we conclude
\begin{eqnarray}\label{F9a}
U(2\pi,0)&=&U(2\pi,\pi)\,U(\pi,0)\\
\label{F9b}
&\stackrel{(\ref{F8a})}{=}&\left({\mathcal T}\,U(\pi,0)\,{\mathcal T}\right)\,U(\pi,0)\\
\label{F9c}
&=&\left( {\mathcal T}\,U(\pi,0)\right)^2
\;.
\end{eqnarray}
Hence the monodromy matrix $U(2\pi,0)$ is completely determined by the ``half-period" monodromy $U(\pi,0)$.
This result resembles an argument in the appendix of \cite{H16} used to reduce the effort of the numerical computation by a factor of two.
Especially the quasienergies can be obtained as twice the argument of the eigenvalues of ${\mathcal T}\,U(\pi,0)$ divided by the period $2\pi$.
The latter is the matrix $U(\pi,0)$ with transposed rows.

Second, according to (\ref{F6c}) the Schr\"odinger equation (\ref{F2}) is also invariant under the combined operation
\begin{equation}\label{F10}
\widetilde{\mathcal C}:
\left\{\begin{array}{c}
         \tau\mapsto \pi-\tau \\
         \mbox{complex conjugation }
       \end{array}
\right.
\;,
\end{equation}
since after the time reflection $\tau\mapsto \pi-\tau$ the system feels the same magnetic field as before.
This invariance implies
\begin{equation}\label{F11}
U(0,\pi)=\overline{U(\pi,0)}
\;.
\end{equation}
On the other hand, $U(0,\pi)$ is the inverse (adjoint) of $U(\pi,0)$, hence both matrices must be symmetric.
Taking into account the special form (\ref{F6}) of evolution matrices for two-level systems, it follows that
the anti-diagonal elements of $U(\pi,0)$ must be purely imaginary, i.~e.~,
\begin{equation}\label{F12}
  U(\pi,0)_{12}= U(\pi,0)_{21}={\sf i}\,\,r,\quad r\in [-1,1]
  \;,
\end{equation}
and hence  $U(\pi,0)$ has the form
\begin{equation}\label{F13}
 U(\pi,0)=\left(
  \begin{array}{cc}
  \sqrt{1-r^2}\,e^{{\sf i}\alpha}& {\sf i}\,r \\
    {\sf i}\,r &\sqrt{1-r^2}\,e^{-{\sf i}\alpha}
  \end{array}
  \right)
  \;,
\end{equation}
with some phase factor $e^{{\sf i}\alpha},\;\alpha\in [0,2\pi)$.
The eigenvalues $\delta_\pm$ of
\begin{equation}\label{F14}
{\mathcal T}\, U(\pi,0)=\left(
  \begin{array}{cc}
      {\sf i}\,r &\sqrt{1-r^2}\,e^{-{\sf i}\alpha}\\
     \sqrt{1-r^2}\,e^{{\sf i}\alpha}& {\sf i}\,r
  \end{array}
  \right)
  \end{equation}
are  $\delta_\pm= {\sf i} \,r\,\pm \sqrt{1-r^2}$, independent of $\alpha$.
From the above considerations it follows that the quasienergies are
\begin{equation}\label{F15}
 \epsilon_\pm=\pm\frac{1}{\pi}\arcsin r
 \;.
\end{equation}
Hence it suffices to know the function $r=r(f,\nu)$ in order to calculate the quasienergies.
Note further that (\ref{F9c}) and (\ref{F14}) yield the following form of the monodromy matrix
\begin{equation}\label{F15}
U(2\pi,0)=\left(
  \begin{array}{cc}
     1-2r^2& 2\,{\sf i}\,r \sqrt{1-r^2}\,e^{-{\sf i}\,\alpha}\\
      2\,{\sf i}\,r \sqrt{1-r^2}\,e^{{\sf i}\,\alpha}&1-2r^2
  \end{array}
  \right)
  \;.
  \end{equation}

A further consequence of the invariance (\ref{F10}) is the following equation:
\begin{equation}\label{F16}
 U(\frac{\pi}{2},0)=\overline{U(\frac{\pi}{2},\pi)}=\overline{U(\pi,\frac{\pi}{2})^{-1}}=U(\pi,\frac{\pi}{2})^{\top}
 \;,
 \end{equation}
where $^\top$ denotes the transposed matrix. It follows that the half-period monodromy matrix can already be obtained from
the ``quarter-period" monodromy by
\begin{equation}\label{F17}
  U(\pi,0)=U(\pi,\frac{\pi}{2})\,U(\frac{\pi}{2},0)\stackrel{(\ref{F16})}{=}U(\frac{\pi}{2},0)^\top\, U(\frac{\pi}{2},0)
  \;.
\end{equation}
 Let us write
 \begin{equation}\label{F18}
   U(\frac{\pi}{2},0)=\left(
  \begin{array}{cc}
    a & -\overline{b}\\
     b&\overline{a}
  \end{array}
  \right)
  \;,
 \end{equation}
 where $a\equiv\psi_1(\frac{\pi}{2})$ and $b\equiv\psi_2(\frac{\pi}{2})$. Then it follows that
  \begin{eqnarray}\label{F19a}
  U(\frac{\pi}{2},0)^\top\, U(\frac{\pi}{2},0)&=&\left(
  \begin{array}{cc}
    a & b\\
    -\overline{b}&\overline{a}
  \end{array}
  \right)\,
  \left(
  \begin{array}{cc}
    a & -\overline{b}\\
     b&\overline{a}
  \end{array}
  \right)
  =
   \left(
  \begin{array}{cc}
    a^2+b^2 & -a \overline{b}+b \overline{a}\\
     -\overline{b}a+\overline{a}b&\overline{a}^2+\overline{b}^2
     \end{array}
  \right)
   \\
     \label{F19b}
     &\stackrel{(\ref{F17})(\ref{F13})}{=}&
         \left(
  \begin{array}{cc}
  \sqrt{1-r^2}\,e^{{\sf i}\alpha}& {\sf i}\,r \\
    {\sf i}\,r &\sqrt{1-r^2}\,e^{-{\sf i}\alpha}
  \end{array}
  \right)
    \;.
 \end{eqnarray}
 Comparison of the matrix elements of (\ref{F19a}) and (\ref{F19b}) yields the auxiliary quantities $r$ and $\alpha$ in terms of the
  quarter period data:
 \begin{eqnarray}
\label{F20a}
  r &=& 2\,\mbox{Im}\left(\overline{a}\,b \right)= 2\,\mbox{Im}\left(\overline{\psi_1\left(\frac{\pi}{2}\right)}\,\psi_2\left(\frac{\pi}{2}\right)\right)\;,\\
  \label{F20b}
  \alpha &=&\arg\left(a^2+b^2 \right)=\arg\left(\psi_1\left(\frac{\pi}{2}\right)^2+\psi_2\left(\frac{\pi}{2}\right)^2 \right)
  \;.
 \end{eqnarray}

 Finally we will show that the time evolution in the first half-period can be reduced to the first quarter-period.
 In fact, the invariance of the Schr\"odinger equation under (\ref{F10}) implies
 \begin{equation}\label{F21}
  U\left(\frac{\pi}{2}+\tau,\frac{\pi}{2}\right)=\overline{ U\left(\frac{\pi}{2}-\tau,\frac{\pi}{2}\right)}=U\left(\frac{\pi}{2},\frac{\pi}{2}-\tau\right)^\top
  \;.
 \end{equation}
 Further we conclude
 \begin{eqnarray}
\label{F22a}
   U\left(\frac{\pi}{2}+\tau,0\right) &=&  U\left(\frac{\pi}{2}+\tau,\frac{\pi}{2}\right)\, U\left(\frac{\pi}{2},0\right) \\
   \label{F22b}
    &\stackrel{(\ref{F21})}{=}& U\left(\frac{\pi}{2},\frac{\pi}{2}-\tau\right)^\top\, U\left(\frac{\pi}{2},0\right)\\
    \label{F22c}
    &=& \left(U\left(\frac{\pi}{2}, 0\right)\,U\left(0, \frac{\pi}{2}-\tau\right) \right)^\top\, U\left(\frac{\pi}{2},0\right)\\
     \label{F22d}
    &=& U\left(0, \frac{\pi}{2}-\tau\right)^\top\,U\left(\frac{\pi}{2}, 0\right)^\top\, U\left(\frac{\pi}{2},0\right)\\
     \label{F22e}
     &=& \overline{U\left( \frac{\pi}{2}-\tau,0\right)}\,U\left(\frac{\pi}{2}, 0\right)^\top\, U\left(\frac{\pi}{2},0\right)\\
      \label{F22f}
     &\stackrel{(\ref{F17})}{=}& \overline{U\left( \frac{\pi}{2}-\tau,0\right)}\,U\left(\pi,0\right)
     \;.
 \end{eqnarray}

 Together with (\ref{F8}) this means that the time evolution of the RPL can be completely reduced to the time evolution
 in the first quarter-period. This is important for the following sections since the calculation of confluent Heun functions corresponding
 to the time evolution in the first quarter-period is especially simple.

\section{Schr\"odinger equation and confluent Heun equation (CHE)}\label{sec:M}

Following \cite{ML07,XieHai10} we will transform the Schr\"odinger equation (\ref{F2}) in the following way.
First, we consider the second derivative of $\psi_1(\tau)$ and, after eliminating $\psi_2(\tau)$, obtain
\begin{equation}\label{T1}
  \frac{d^2}{d\tau^2}\psi_1(\tau)+\left(
  {\sf i}\,\frac{f}{2}\,\cos\tau +\frac{1}{4}\,f^2 \,\sin^2\tau+\frac{1}{4}\,\nu^2
  \right)\psi_1(\tau)=0
  \;.
\end{equation}

\noindent $\psi_2(\tau)$ satisfies a similar second order equation that need not be considered here. Passing to a second order equation
enlarges the solution space, but this is harmless as far as the initial conditions for (\ref{T1}) are chosen according to the
first order Schr\"odinger equation (\ref{F2}). Next we consider the transformation
\begin{equation}\label{T2}
z(\tau)=\sin^2\frac{\tau}{2}=\frac{1}{2}\left( 1-\cos\tau\right)
\;,
\end{equation}
restricted to a bijective $C^\infty$--map $z:[0,\pi]\rightarrow [0,1]$, and the function $y:[0,1]\rightarrow{\mathbbm C}$ defined by
\begin{equation}\label{T3}
 y(z(\tau))=\exp\left(-{\sf i}\,f\,z(\tau) \right) \psi_1(\tau)
 \;.
\end{equation}
It is straightforward to transform (\ref{T1}) into a linear second order differential equation for $y(z)$:
\begin{eqnarray}
\nonumber
0&=&\frac{d^2}{d z^2} y(z) +\left(\frac{1}{2z} +\frac{1}{2(z-1)}+2\,{\sf i}\, f\right) \frac{d}{d z}y(z)
+\left({\sf i}\, f \,(2 z-1)-\frac{\nu ^2}{4}
\right)\frac{y(z)}{z(z-1)}.\\
\label{T4}
&&
\end{eqnarray}
It has the form of a confluent Heun equation (CHE), see, e.~g., \cite{DLMF}, 31.12.1, 
\begin{eqnarray}
\nonumber
0&=&\frac{d^2}{d z^2} y(z) +\left(\frac{1-\mu_0}{z} +\frac{1-\mu_1}{z-1}+a\right) \frac{d}{d z}y(z)\\
\nonumber
&+&\left( \frac{1}{2}(1-\mu_0)(1-\mu_1)+\frac{a}{2}\left( (1-\mu_0)(z-1)+(1-\mu_1)z\right)+b_0+b_1 z
\right)\frac{y(z)}{z(z-1)},\\
\label{T5}
&&
\end{eqnarray}
where the five complex parameters $\mu_0,\mu_1,a,b_0,b_1$ are functions of the two physical parameters $f,\nu$:
\begin{eqnarray}
\label{T6a}
  \mu_0 &=& \mu_1=\frac{1}{2} \\
  \label{T6b}
  a &=& 2\,{\sf i}\,f \\
  \label{T6c}
  b_0 &=& -\frac{1}{8} \left(4\, {\sf i}\, f+2\, \nu ^2+1\right) \\
  \label{T6d}
  b_1 &=& {\sf i}\,f
  \;.
\end{eqnarray}
Here and in what follows we will stick closely to the notation of \cite{SS80} in order to facilitate the comparison of equations.
Usually, the dependence on the five parameters will be suppressed with the exception of ${\boldsymbol \mu}\equiv (\mu_0,\mu_1)$.
The CHE has two regular singular points at $z=0$ and $z=1$ (and an irregular singular point at $z=\infty$). It is possible to
devise power series solutions around $z=0$ and $z=1$ that, following \cite{SS80} (except for a factor), will be written as:
\begin{equation}\label{T7}
  \eta_0(z,{\boldsymbol \mu})=\sum_{k=0}^\infty  \tau_k^0({\boldsymbol \mu})\,z^k
  \;,
\end{equation}
and
\begin{equation}\label{T8}
  \eta_1(z,{\boldsymbol \mu})=\sum_{k=0}^\infty  \tau_k^1({\boldsymbol \mu})\,(1-z)^k
  \;.
\end{equation}
The radius of convergence of both series is ${\sf R}=1$.
We  set $\tau_0^0=1$ which implies
\begin{equation}\label{T9a}
\eta_0(0,{\boldsymbol \mu})=1
\;.
\end{equation}

Both sets of coefficients satisfy three-term recurrence relations, see \cite{S06} or  \cite{XieHai10}, together with the initial values $\tau_{-1}^0=0$ and $\tau_0^0=1$.
Here we explicitly only mention
those for $\tau_k^0({\boldsymbol \mu})$ where ${\boldsymbol \mu}=(\frac{1}{2},\frac{1}{2})$ or ${\boldsymbol \mu}=(-\frac{1}{2},\frac{1}{2})$
since the other ones are not needed in this paper:
\begin{eqnarray}
\label{T10a}
\boldsymbol{\mu} =\left(\frac{1}{2},\frac{1}{2}\right):\;\tau_{k+1}^0 &=& \frac{-4 {\sf i} f (2 k+1)+4 k^2-\nu ^2}{2(k+1) (2   k+1)} \tau_k^0
+\frac{8 {\sf i} f k}{2 (k+1) (2k+1)}\tau_{k-1}^0,\\
\label{T10b}
\boldsymbol{\mu} =\left(-\frac{1}{2},\frac{1}{2}\right):\;\tau_{k+1}^0 &=& \frac{4 (k+1) (k-2 {\sf i} f)-\nu ^2+1}{2 (k+1) (2 k+3)} \tau_k^0
+\frac{2{\sf i} f(2 k+1)}{(k+1) (2 k+3)}\tau_{k-1}^0
\;.
\end{eqnarray}
In particular,
\begin{eqnarray}
\label{T10c}
  \tau_{1}^0\left(\frac{1}{2},\frac{1}{2}\right) &=& -\frac{\nu ^2}{2}-2\,{\sf i}\, f
  \;.
 \end{eqnarray}

It is crucial to distinguish between global solutions of the CHE (\ref{T5}) and local series representations as (\ref{T7}) and (\ref{T8}).
In order to simplify the presentation $\eta_n(z,{\boldsymbol \mu}),\;n=0,1$ will also denote appropriate analytical continuations of the corresponding series representations  (\ref{T7}) and (\ref{T8}).
According to \cite{SS80} there exist two fundamental systems of global solutions, denoted by $(y_{01},y_{02})$ and $(y_{11},y_{12})$,
that are holomorphic at least in the common domain
\begin{equation}\label{G1}
 {\mathcal H}\equiv {\mathbbm C}\backslash \left( (-\infty,0] \cup [1,\infty)\right)
 \;,
 \end{equation}
i.~e.~, in the complex plane with two cuts ending at the singular points $z=0$ and $z=1$.
We remark that the authors of \cite{SS80} consider a smaller domain since they assume a more general differential equation than the CHE that might have additional
singular points.
However, for the CHE the domain ${\mathcal H}$ is appropriate, see \cite{S06}, where ${\mathcal H}$ is further extended into a Riemannian surface.
The fundamental systems are defined by:
\begin{eqnarray}
\label{G2a}
y_{01}&\equiv&\eta_0(z,\mu_0,\mu_1),\\
\label{G2b}
y_{02}&\equiv& z^{\mu_0}\,\eta_0(z,-\mu_0,\mu_1),\\
\label{G2c}
y_{11}&\equiv&\eta_1(z,\mu_0,\mu_1),\\
\label{G2d}
y_{12}&\equiv& (1-z)^{\mu_1}\,\eta_1(z,\mu_0,-\mu_1)\;.
\end{eqnarray}
Hence $y_{01}$ is also holomorphic in the open unit disc with center $z=0$ that exceeds ${\mathcal H}$, analogously $y_{11}$ for $z=1$.
In contrast, $y_{02}$ is the product of a holomorphic function with, in our case, the factor $\sqrt{z}$ and hence has a branch point
at $z=0$, analogously for $y_{12}$ at $z=1$.

Since we have reduced the time evolution of the RPL to the first quarter-period in section \ref{sec:F} which corresponds to the interval
$z\in[0,\frac{1}{2}]$ for the arguments of the Heun functions, it will suffice to consider the first fundamental system
$(y_{01},y_{02})$
that can be expressed through the Heun functions $\eta_0(z,{\boldsymbol \mu})$
given by the series representation (\ref{T7}). We will not need to switch to $(y_{11},y_{12})$ and hence need not consider
the corresponding connection problem, see \cite{SS80,S06}.

\section{Calculation of the quasienergy }\label{sec:C}

We reconsider the solution $\psi(\tau)$ of the Schr\"odinger equation (\ref{F2}) subject to the initial condition
$\psi_1(0)=1$ and $\psi_2(0)=0$. Let $Y(z(t))=\exp(-{\sf i}f z((\tau))\psi_1(\tau)$ be the corresponding solution of the CHE according to (\ref{T3}).
It satisfies $Y(0)=1$ and its derivative is given by
\begin{equation}\label{C1}
  \frac{dY}{dz}=-{\sf i}\,f\,Y(z)+\exp(-{\sf i}\,f\, z)\,\frac{\frac{d\psi_1}{d\tau}}{\frac{dz}{d\tau}}
  \;.
\end{equation}
For $\tau\rightarrow 0$ both expressions $\frac{d\psi_1}{d\tau}=-{\sf i}\frac{f}{2}\sin\tau\;\psi_1-{\sf i}\frac{\nu}{2}\psi_2$ and $\frac{dz}{d\tau}=\frac{1}{2}\sin\tau$ vanish
and we have to calculate the limit of (\ref{C1}) by L'Hospital's rule. After some elementary calculations we thereby obtain
\begin{equation}\label{C2}
  \frac{dY(0)}{dz}=-2\,{\sf i}\,f-\frac{\nu^2}{2}
  \;.
\end{equation}
Comparison with (\ref{T10c}) yields
\begin{equation}\label{C2a}
 Y(z)=\eta_0\left(z,\frac{1}{2},\frac{1}{2}\right)=y_{01}(z)
 \;,
\end{equation}
and hence
\begin{equation}\label{C2b}
 \psi_1(\tau)=\exp({\sf i}f z(\tau)\,Y(z(t))=\exp\left({\sf i}f \sin^2(\tau/2)\right)\,\eta_0\left(\sin^2(\tau/2),\frac{1}{2},\frac{1}{2}\right)
 \;,
\end{equation}
for $t\in[0,\pi)$ according to the convergence domain of the power series for $\eta_0$.

Next we consider $U(\tau,0)_{21}=\psi_2(\tau)$, see (\ref{F6}), and will determine its initial conditions. By assumption,
$\psi_2(0)=0$ and due to the Schr\"odinger equation (\ref{F2}),
$\frac{d\psi_2(0)}{d\tau}=-{\sf i}\,\frac{\nu}{2}\psi_1(0)=-{\sf i}\,\frac{\nu}{2}$.
Recall that the second column of $U(\tau,0)$ is a second solution $\widetilde{\psi}(\tau)$ of (\ref{F2})
that is determined by $\psi(\tau)$ via (\ref{F6}) and hence satisfies
\begin{equation}\label{C3}
 \widetilde{\psi}_1=-\overline{\psi_2}
 \;:
\end{equation}
This implies that $\widetilde{\psi}(\tau)$  has the initial values
$\widetilde{\psi}_1(0)=-\overline{\psi_2}(0)=0$
and
\begin{equation}\label{CLimit}
 \frac{d\widetilde{\psi_1}(0)}{d\tau}=-\frac{d\overline{\psi_2}(0)}{d\tau}=-{\sf i}\,\frac{\nu}{2}
 \;.
\end{equation}
Let $Z(z(\tau))=\exp(-{\sf i}f z(\tau))\widetilde{\psi_1}(\tau)$ be the corresponding solution of the CHE according to (\ref{T3}).
It satisfies $Z(0)=0$ and its derivative is given by
\begin{equation}\label{C5}
  \frac{dZ}{dz}=-{\sf i}\,f\,Z(z)+\exp(-{\sf i}\,f\, z)\,\frac{\frac{d\widetilde{\psi_1}}{d\tau}}{\frac{dz}{d\tau}}
  \;.
\end{equation}
In the limit $\tau\rightarrow 0$ the derivative $\frac{dZ}{dz}$ diverges since $\frac{d\widetilde{\psi_1}}{d\tau}$ assumes a finite
value but $\frac{dz}{d\tau}=\frac{1}{2}\sin\tau$ vanishes.
Being a solution of the CHE, $Z$ must be a linear combination of the fundamental system $(y_{01},y_{02})$, $Z=\lambda_1\,y_{01}+\lambda_2\,y_{02}$.
It follows that  $0=Z(0)=\lambda_1\,y_{01}(0)+\lambda_2\,y_{02}(0)=\lambda_1$, using (\ref{G2a}), (\ref{G2b}) and (\ref{T9a}),
and hence $\lambda_1=0$.
This in turn implies
$\widetilde{\psi_1}(\tau)=\lambda_2\,\exp({\sf i}f z)\,y_{02}(z)=\lambda_2\,\exp({\sf i}f z)\,\sqrt{z}\,\eta_0\left(z,-\frac{1}{2},\frac{1}{2}\right)$.
To determine $\lambda_2$ we consider
\begin{eqnarray}
\label{C6a}
  \frac{d\widetilde{\psi_1}}{d\tau} &=&\frac{dz}{d\tau}\frac{d}{dz}\left(\lambda_2\,\exp({\sf i}f z)\,\sqrt{z}\,\eta_0\left(z,-\frac{1}{2},\frac{1}{2}\right)\right)\\
  \label{C6b}
  &=&\frac{1}{2}\sin\tau\left({\sf i}\,f\,\widetilde{\psi_1}(\tau)+\lambda_2\,\exp({\sf i}f z)\,\frac{1}{2\sqrt{z}}\,
  \eta_0\left(z,-\frac{1}{2},\frac{1}{2}\right)+\lambda_2\,\exp({\sf i}f z)\,\sqrt{z}\,\frac{d\eta_0\left(z,-\frac{1}{2},\frac{1}{2}\right)}{dz}\right).
 \end{eqnarray}
In the limit $\tau\rightarrow 0$ only the second last term of (\ref{C6b}) survives and yields
\begin{eqnarray}\nonumber
 \lim_{\tau\rightarrow 0} \frac{d\widetilde{\psi_1}}{d\tau}
 &=&\lim_{\tau\rightarrow 0}\frac{1}{2}\,\lambda_2\,\exp({\sf i}f z(\tau))\,\cos\frac{\tau}{2}\,\eta_0\left(z(\tau),-\frac{1}{2},\frac{1}{2}\right)\\
 \label{C7}
 &\stackrel{(\ref{T9a})}{=}&\frac{\lambda_2}{2}
 \;.
\end{eqnarray}
Comparison with (\ref{CLimit}) yields
\begin{equation}\label{C8}
 \lambda_2=-{\sf i}\,\nu
 \;,
\end{equation}
and hence
\begin{equation}\label{C9}
  \widetilde{\psi_1}(\tau)=-{\sf i}\,\nu \,\exp({\sf i}f z(\tau))\,\eta_0\left(z(\tau),-\frac{1}{2},\frac{1}{2}\right)
  \;,
\end{equation}
or, according to (\ref{C3}),
\begin{equation}\label{C10}
  \psi_2(\tau)=-{\sf i}\,\nu \,\exp(-{\sf i}f \sin^2(\tau/2))\,\overline{\eta_0\left(\sin^2(\tau/2),-\frac{1}{2},\frac{1}{2}\right)}
  \;,
\end{equation}
for $\tau\in [0,\pi)$.

Next, we choose the special value $\tau=\frac{\pi}{2}$ corresponding to $z(\tau)=\frac{1}{2}$ and insert (\ref{C2b}) and (\ref{C10}) into
(\ref{F20a}) and (\ref{F20b}). Further we will use the following abbreviations
\begin{eqnarray}
\label{C10a}
  \eta_{++} &\equiv& \left.\eta_0\left(z,\frac{1}{2},\frac{1}{2}\right)\right|_{z=\frac{1}{2}}, \\
  \label{C10b}
  \eta_{-+} &\equiv& \left.\eta_0\left(z,-\frac{1}{2},\frac{1}{2}\right)\right|_{z=\frac{1}{2}}
  \;,
\end{eqnarray}
where the dependence on the physical parameters $f$ and $\nu$ is usually suppressed.
 After some elementary transformations we then obtain the following expressions for the auxiliary quantities
\begin{eqnarray}
\label{C11a}
 r &=& -\sqrt{2}\,\nu\,\mbox{Re}
 \left( e^{{\sf i}f}\, \eta_{++}\, \eta_{-+}\right), \\
 \label{C11b}
 \alpha &=& \arg\left(
 e^{{\sf i}f}\,\eta_{++}^2 -
 \frac{\nu^2}{2}\,e^{-{\sf i}f}\,\overline{\eta_{-+}}^2
 \right)
 \;.
\end{eqnarray}

In view of (\ref{F15}) this yields the following explicit expression for the dimensionless quasienergies
\begin{equation}\label{C12}
 \pm\epsilon(f,\nu)= \mp\,\frac{1}{\pi}\arcsin \left(
  \sqrt{2}\,\nu\,\mbox{Re}
 \left( e^{{\sf i}f}\,\eta_{++}(f,\nu)\,\eta_{-+}(f,\nu)\right)
  \right)
  \;.
\end{equation}

\begin{figure}[h]
  \centering
    \includegraphics[width=1.0\linewidth]{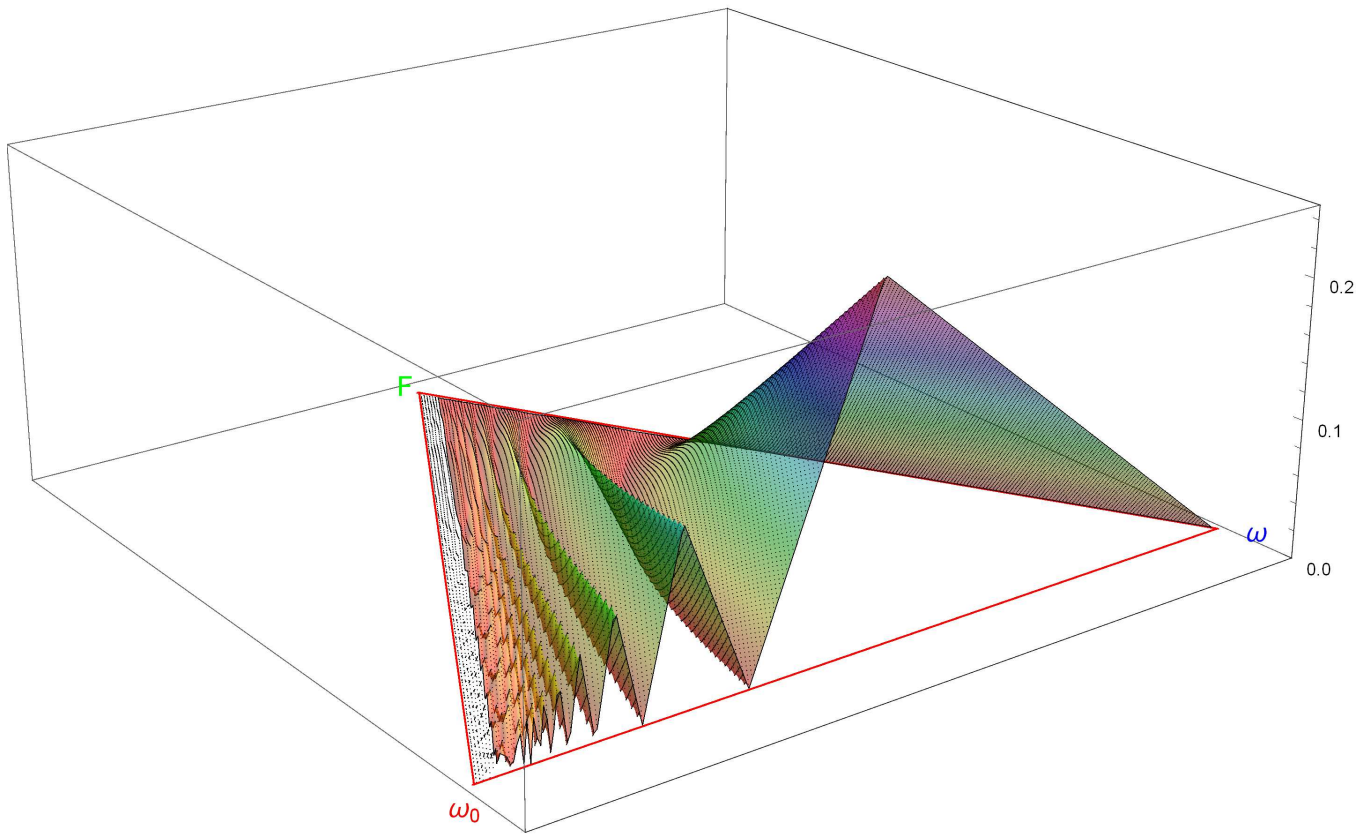}
  \caption[EPSH]
  {The quasienergy ${\mathcal E}$ as a function of the scaled variables $\omega_0,\,\omega$ and $F$ subject to the constraint $\omega_0+\omega+F=1$.
  At the vertices of the basic triangle the marked variable has the value $1$ and the remaining two variables vanish.
  The black dots are calculated by numerically solving the Schr\"odinger equation (\ref{F2}). The colored graph of ${\mathcal E}$ is calculated by using the analytical form (\ref{C12}).
  }
  \label{FIGEPSI}
\end{figure}

In Figure \ref{FIGEPSI} we have plotted the quasienergy ${\mathcal E}$, see (\ref{F3a}),
as a function of the three scaled positive variables $\omega_0,\,\omega$ and $F$ subject to the constraint $\omega_0+\omega+F=1$ that can be represented by the points of an equilateral triangle, see \cite{S18}. We will shortly explain this representation. The domain of arguments $\omega_0,\,\omega$ and $F$ of the quasienergy
${\mathcal E}$ is the positive octant $P$ of ${\mathbbm R}^3$ and a representation of the graph of ${\mathcal E}$ would be impossible in three dimensions. But we can
exploit the fact that ${\mathcal E}$ is a positively homogeneous function, i.~e.~,
\begin{equation}\label{Q9a}
 {\mathcal E}(\lambda\, \omega_0,\lambda\,\omega,\lambda\,F)=\lambda\, {\mathcal E}( \omega_0,\omega,F)
\end{equation}
for all $\lambda>0$. Hence it suffices to represent the graph of ${\mathcal E}$ for a two-dimensional section of $P$. As such a section we choose
the intersection of $P$ with the plane defined by $\omega_0+\omega+F=1$, which is just the equilateral triangle mentioned above.
It turns out that the series representation of the confluent Heun functions that enter into (\ref{C12}) is not sufficiently accurate if the values of
$\nu=\frac{\omega_0}{\omega}$ and $f=\frac{F}{\omega}$ are too large, that means for too small $\omega$.
This cannot be fixed by increasing the number of terms used to approximate the series.
Hence we  proceeded as follows:
First we have divided the equilateral triangle of scaled variables uniformly into
smaller triangles and chosen $32,385$ points in the interior where the quasienery has been calculated by numerically solving the Schr\"odinger
equation. Then we have chosen a subset  of $30,876$ points where the scaled frequency satisfies $\omega>3/128$. For the points of this
subset the quasienergy has been calculated by using the exact formula (\ref{C12}) and $N=100$ terms of the two series involved.
It turns out that the maximal deviation between the two values of the quasienergy calculated as described is smaller than $1.3\times 10^{-4}$.
Hence this deviation is not visible in Figure  \ref{FIGEPSI}. However, this result shows that it might be advantageous to use analytical approximations
for the quasienergy that are valid in the adiabatic limit $\omega\rightarrow 0$.

\begin{figure}[h]
  \centering
    \includegraphics[width=0.75\linewidth]{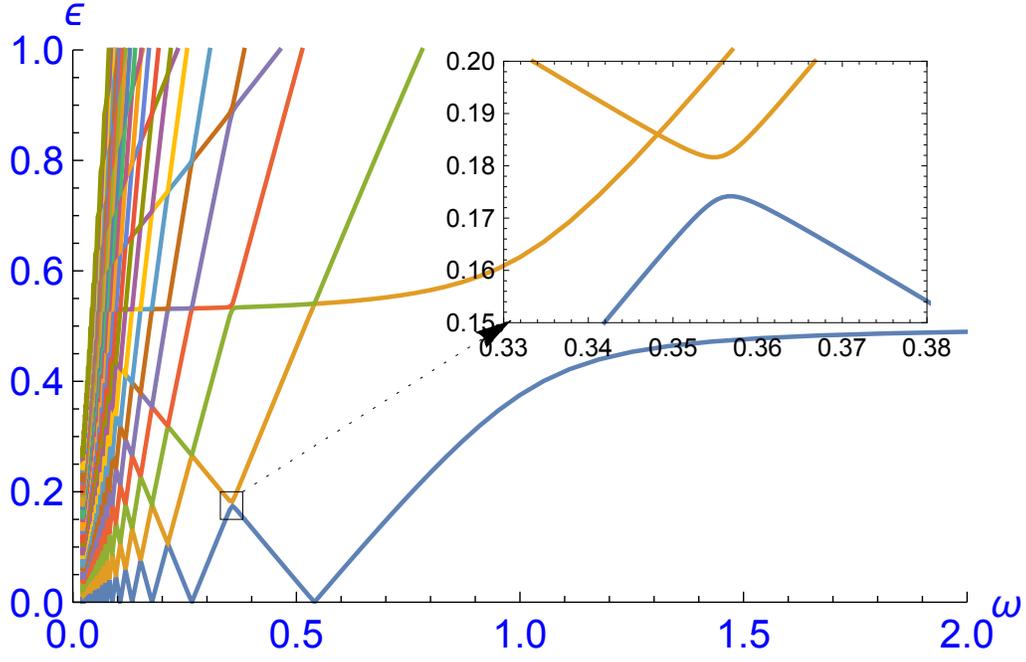}
  \caption[EP]
  {Various branches of the quasienergy derived from ${\mathcal E}(\omega)$ according to (\ref{Q9b}) where we have chosen $F=1/2$ and $\omega_0=1$.
   The blue curves have been calculated by using (\ref{C12}) and a series truncation of $N=100$ terms.
  Then, for example, the dark yellow curves are obtained from the blue ones by adding the linear function
  $\omega$ to $-{\mathcal E}(\omega)$, the green curves by $\omega+{\mathcal E}(\omega)$, etc.~.  The inset demonstrates the avoided level crossing
  between two branches in the neighbourhood of the resonance frequency $\omega_{res}^{(2)}\approx 0.355776$ that is hardly visible in the original graphics.
  }
  \label{FIGLEP}
\end{figure}

Finally we note that the function ${\mathcal E}( \omega_0,\omega,F)$ gives rise to an infinite variety of derived quasienergy branches of the form
\begin{equation}\label{Q9b}
 \pm\, {\mathcal E}(\omega_0,\omega,F)+n\,\hbar\,\omega,\quad n\in{\mathbbm Z}
 \;,
\end{equation}
see Figure \ref{FIGLEP}. This corresponds to the choice of different branches of the $\arcsin$-function in (\ref{C12}) for even $n$.
The cases of odd $n$ are obtained as a consequence of the equation $\sin(\pi\,\epsilon+\pi)=\sin(-\pi\,\epsilon)$.

\vspace{5mm}

\begin{figure}[h]
  \centering
    \includegraphics[width=0.75\linewidth]{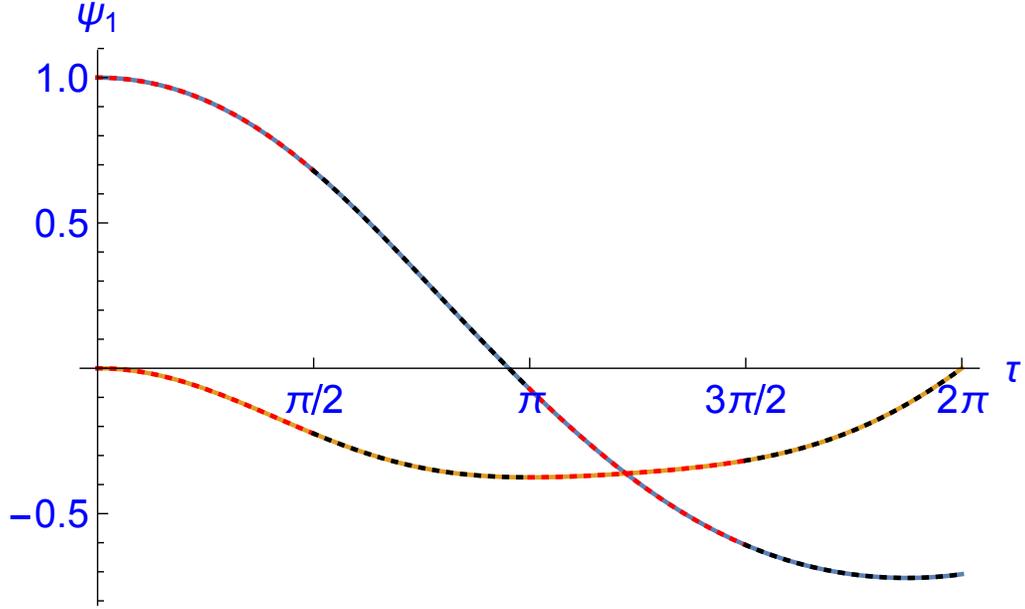}
  \caption[T1]
  {The first component $\psi_1(\tau)= u_1(\tau)+{\sf i}\,v_1(\tau)$ of the solution (\ref{TT3}) of the Schr\"odinger equation for $f=1/2$, $\nu=1$
  and $\tau\in [0,2\pi]$ calculated by different methods.
  The blue curve shows $u_1(\tau)$ and the dark yellow one $v_1(\tau)$, obtained by numerically solving the Schr\"odinger equation.
  The dotted curves are calculated by using two series solutions of the CHE for $0\le z\le 1/2$ with $1,000$ terms and
  the equations (\ref{TT2a}) -- (\ref{TT4d}) that reduce the time evolution to the first quarter-period.
  Note that $v_1(2\pi)=0$ since $\psi_1(2\pi)$ is real according to (\ref{F15}).
   }
  \label{FIGT1}
\end{figure}

\begin{figure}[h]
  \centering
    \includegraphics[width=0.75\linewidth]{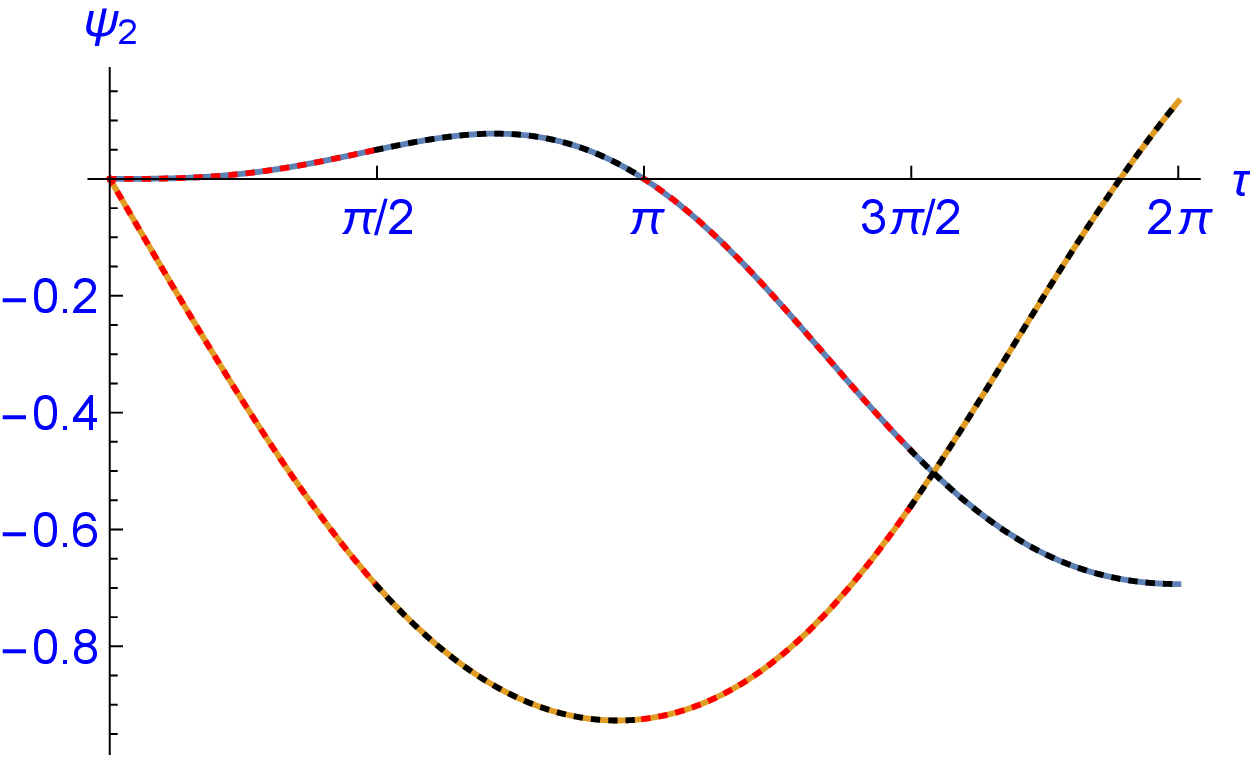}
  \caption[T2]
  {The second component $\psi_2(\tau)= u_2(\tau)+{\sf i}\,v_2(\tau)$ of the solution (\ref{TT3}) of the Schr\"odinger equation analogous to Figure \ref{FIGT1}.
 Note that $u_2(\pi)=0$ since $\psi_2(\pi)$ is purely imaginary according to (\ref{F12}).
    }
  \label{FIGT2}
\end{figure}

\subsection{The limit of the quasienergy for $\omega_0\rightarrow 0$ }\label{sec:L}

In the limit $\omega_0 \to 0$ the Schr\"odinger equation can be solved in
linear order with respect to $\omega_0$. The quasienergy then is obtained
as
\begin{equation}\label{L1}
{\mathcal E} = \frac{\hbar\,\omega_0}{2}\,J_0\left( \frac{F}{\omega}\right)+O(\omega_0^3)
  \;,
\end{equation}
where $J_0$ denotes the Bessel function of zeroth order; see, for example,
Ref.~\cite{S18} for a systematic derivation. This approximation, which was
known already to Shirley (see Eq.~(27) in Ref.~\cite{Shirley65}) is of
substantial practical importance. It explains, among other things, the
effective quenching of the level splitting of the dressed ground-state doublet
in a periodically driven symmetric double well, and can easily be generalized
to describe the narrowing of tightly bound Bloch bands in lattice potentials
under the action of strong time-periodic forcing~\cite{Holthaus92}. The latter
effect has been exploited recently in a number of experiments performed with
ultracold atoms in periodically driven optical lattices~\cite{ZenesiniEtAl09,
StruckEtAl11,AidelsburgerEtAl13,MiyakeEtAl13,Eckardt17,GorgEtAl18}, having
become a key instrument of Floquet engineering. Therefore, it is of particular
interest to derive this result directly from the explicit representation (\ref{C12}).

To this end we have to calculate the values of $\eta_{++}(f,\nu)$ and $\eta_{-+}(f,\nu)$ that occur in (\ref{C12})
in the limit $\nu\rightarrow 0$. For $\nu=0$ the CHE (\ref{T4}) assumes the form
\begin{equation}\label{L2}
 0=\frac{d^2}{d z^2} y(z) +\left(\frac{1}{2z} +\frac{1}{2(z-1)}+2\,{\sf i}\, f\right) \frac{d}{d z}y(z)
+{\sf i}\, f \,(2 z-1)\frac{y(z)}{z(z-1)}
\;.
\end{equation}
This differential equation admits the special solution $Y(z)=e^{-2\,{\sf i}\,f\,z}$ with the initial conditions $Y(0)=1$
and $\frac{dY(0)}{dz}=-2\,{\sf i} f$. Comparison with (\ref{C2}) shows that $Y$ is the limit $\nu\rightarrow 0$ of the first fundamental solution
$y_{01}$ of the CHE  and hence
\begin{equation}\label{L3}
\eta_{++}(f,0)=Y\left(\frac{1}{2}\right)=e^{-{\sf i}\,f}
\;.
\end{equation}
In order to calculate $\eta_{-+}(f,0)$ we transform the differential equation (\ref{L2}) according to
\begin{equation}\label{L4}
 y(z) = e^{-2\,{\sf i}\,f\,z}\, w(z)
\end{equation}
into
\begin{equation}\label{L5}
  0=2 (z-1) z w''(z)+\left(-4 \,{\sf i} f z^2+(2+4 {\sf i} f) z-1\right) w'(z)
  \;.
\end{equation}
By separation of variables we obtain its general solution
\begin{equation}\label{L6}
w(z)= A+B \int \frac{e^{2 \,{\sf i} f z}}{\sqrt{(1-z) z}} \, dz
\;,
\end{equation}
with integration constants $A,B$. The choice $A=1,\, B=0$  reproduces the above solution $Y(z)$.
The choice $A=0,\, B=1$ gives the second fundamental solution of (\ref{L2}). Since we have to evaluate
it at $z=\frac{1}{2}$ we are lead to the integral
\begin{eqnarray}\label{L7a}
I(f)&\equiv&\int_0^{1/2} \frac{e^{2 \,{\sf i} f z}}{\sqrt{(1-z) z}} \, dz\\
\label{L7b}
&=&\sum_{n=0}^\infty \frac{(2\, {\sf i} f )^n}{n!}\,\int_0^{1/2} \frac{z^n}{\sqrt{(1-z) z}} \, dz\\
\label{L7c}
&=&\sum_{n=0}^\infty \frac{(2\,{\sf i} f )^n}{n!}\,B_{\frac{1}{2}}\left(n+\frac{1}{2},\frac{1}{2}\right)\\
\label{L7d}
&=&\frac{1}{2} \pi  e^{{\sf i}\, f} (J_0(f)-{\sf i}\, \pmb{H}_0(f))
\;,
\end{eqnarray}
where $B_a(b,c)$ denotes the incomplete Beta function and  $\pmb{H}_0$ the Struve function of zeroth order.
Resolving the above definitions we obtain
\begin{equation}\label{L8}
\eta_{-+}(f,0)=\frac{e^{-{\sf i} f} }{\sqrt{2}} \, I(f)=\frac{\pi}{2\sqrt{2}} \left(J_0(f)-{\sf i}\, \pmb{H}_0(f)\right)
\;,
\end{equation}
and, finally,
\begin{equation}\label{L9}
\epsilon= \frac{1}{\pi}\arcsin \left(
  \sqrt{2}\,\nu\,\mbox{Re}
 \left( e^{{\sf i}f}\,\eta_{++}(f,0)\,\eta_{-+}(f,0)\right)
  \right)
  \stackrel{(\ref{L3})(\ref{L4})}{=}
  \frac{1}{\pi }\arcsin\left(\frac{1}{2} \pi  \nu  J_0(f)\right)
  = \frac{\nu}{2}J_0(f)+O(\nu^3)
  \;,
\end{equation}
in accordance with (\ref{L1}).

\section{Time evolution}\label{sec:TT}

We will put together the results obtained so far in order to describe a typical time evolution of the RPL.
To this end we will assume that for the physical parameters $f,\,\nu$ under consideration  the values of
the auxiliary quantities $r$ and $\alpha$ have been determined by means of (\ref{C11a}) and (\ref{C11b}).

Recall that, according to (\ref{F8}), the time evolution in the second half-period $[\pi,2\,\pi]$ is completely determined by
the solution of the Schr\"odinger equation (\ref{F2}) for $\tau\in[0,\pi]$. By evaluating
\begin{equation}\label{TT1}
 U(\pi+\tau,0)= U(\pi+\tau,\pi)\, U(\pi,0)\stackrel{(\ref{F8})}{=}{\mathcal T}\,U(\tau,0)\,{\mathcal T}\, U(\pi,0)
 \;,
\end{equation}
and using (\ref{F6}) and (\ref{F13}), we obtain
\begin{eqnarray}
\label{TT2a}
 u_1(\pi+\tau) &=& \sqrt{1-r^2}\left(u_1(\tau) \cos\alpha+v_1(\tau ) \sin \alpha \right) -r\,v_2(\tau)\\
\label{TT2b}
 v_1(\pi+\tau)&=&  \sqrt{1-r^2}\left(-v_1(\tau) \cos\alpha+u_1(\tau ) \sin \alpha \right)+r\,u_2(\tau)\\
 \label{TT2c}
 u_2(\pi+\tau) &=& \sqrt{1-r^2}\left(-u_2(\tau) \cos\alpha-v_2(\tau ) \sin \alpha \right)-r\,v_1(\tau)\\
\label{TT2d}
 v_2(\pi+\tau)&=&  \sqrt{1-r^2}\left(v_2(\tau) \cos\alpha-
 u_2(\tau ) \sin \alpha \right) +r\,u_1(\tau)
 \;,
\end{eqnarray}
where
\begin{equation}\label{TT3}
\psi(\tau)={ u_1(\tau)+{\sf i}\,v_1(\tau)\choose  u_2(\tau)+{\sf i}\,v_2(\tau) }
\end{equation}
is the solution of the Schr\"odinger equation (\ref{F2}) with the initial condition $\psi(0)={1 \choose 0}$.

Analogously, according to (\ref{F22f}), the time evolution in the second quarter-period $[\pi/2,\pi]$ is completely determined by
the solution of the Schr\"odinger equation (\ref{F2}) for $\tau\in[0,\pi/2]$:
\begin{eqnarray}
\label{TT4a}
 u_1\left(\frac{\pi}{2}+\tau\right) &=& \sqrt{1-r^2}\left(u_1\left(\frac{\pi}{2}-\tau\right)  \cos\alpha+v_1\left(\frac{\pi}{2}-\tau\right) \sin \alpha \right) +r\,v_2\left(\frac{\pi}{2}-\tau\right),\\
\label{TT4b}
 v_1\left(\frac{\pi}{2}+\tau\right) &=&  \sqrt{1-r^2}\left(u_1\left(\frac{\pi}{2}-\tau\right)  \sin\alpha-v_1\left(\frac{\pi}{2}-\tau\right) \cos \alpha \right)-r\,u_2\left(\frac{\pi}{2}-\tau\right),\\
 \label{TT4c}
 u_2\left(\frac{\pi}{2}+\tau\right) &=& \sqrt{1-r^2}\left(u_2\left(\frac{\pi}{2}-\tau\right)\cos\alpha+v_2\left(\frac{\pi}{2}-\tau\right) \sin \alpha \right)-r\,v_1\left(\frac{\pi}{2}-\tau\right),\\
\label{TT4d}
 v_2\left(\frac{\pi}{2}+\tau\right)&=&  \sqrt{1-r^2}\left(u_2\left(\frac{\pi}{2}-\tau\right) \sin\alpha-
 v_2\left(\frac{\pi}{2}-\tau\right) \cos \alpha \right) +r\,u_1\left(\frac{\pi}{2}-\tau\right)
 \;.
\end{eqnarray}

Combining the equations (\ref{TT2a}) -- (\ref{TT4d})  the complete time evolution can be reduced to the first quarter-period.
For example, the time evolution in the fourth quarter-period is given by that in the first quarter-period according to the following equations:
\begin{eqnarray}
\label{TT5a}
  u_1\,\left(\frac{3\pi}{2}+\tau\right) &=& (1-2r^2) u_1\left(\frac{\pi}{2}-\tau\right)+2\,r\,\sqrt{1-r^2}\,
  \left(v_2\left(\frac{\pi}{2}-\tau\right)\cos\alpha - u_2\left(\frac{\pi}{2}-\tau\right)\sin\alpha\right), \\
  \label{TT5b}
   v_1\,\left(\frac{3\pi}{2}+\tau\right) &=& (1-2r^2) v_1\left(\frac{\pi}{2}-\tau\right)+2\,r\,\sqrt{1-r^2}\,
  \left(u_2\left(\frac{\pi}{2}-\tau\right)\cos\alpha + v_2\left(\frac{\pi}{2}-\tau\right)\sin\alpha\right), \\
  \label{TT5b}
   u_2\,\left(\frac{3\pi}{2}+\tau\right) &=& -(1-2r^2) u_2\left(\frac{\pi}{2}-\tau\right)+2\,r\,\sqrt{1-r^2}\,
  \left(v_1\left(\frac{\pi}{2}-\tau\right)\cos\alpha - u_1\left(\frac{\pi}{2}-\tau\right)\sin\alpha\right), \\
  \label{TT5b}
   v_2\,\left(\frac{3\pi}{2}+\tau\right) &=& -(1-2r^2) v_2\left(\frac{\pi}{2}-\tau\right)+2\,r\,\sqrt{1-r^2}\,
  \left(u_1\left(\frac{\pi}{2}-\tau\right)\cos\alpha + v_1\left(\frac{\pi}{2}-\tau\right)\sin\alpha\right)
  \;.
\end{eqnarray}

In the first quarter-period we can use the equations (\ref{C2b}) and (\ref{C10}) where the confluent Heun functions $\eta_0\left( z,\frac{1}{2},\frac{1}{2}\right)$
and $\eta_0\left( z,-\frac{1}{2},\frac{1}{2}\right)$ can be evaluated by the corresponding power series without problems (if $\omega$ is not too small, see section \ref{sec:C}).

We have performed such a calculation for the choice of the physical parameters $f=1/2$ and $\nu=1$, where $r=-0.924176\ldots$ and $\alpha=-1.75978\ldots$.
It turns out that an approximation of the two power series involved using $1,000$ terms yields satisfactory results for all values of $z\in(0,1/2)$ when compared with
the direct numerical solution of the Schr\"odinger equation, see Figures \ref{FIGT1} and \ref{FIGT2}. It is not necessary to switch to
the power series (\ref{T8}) about the point $z=1$ by using the corresponding connection equations.

\section{Summary and outlook}\label{sec:SO}

Although the analytical solution \cite{ML07} of the Rabi problem with linear polarization has been published a decade ago
there exist relatively few papers that use this solution. This may be due to the fact that confluent Heun functions are
not so thoroughly investigated as compared with other special functions and that the analytical solution does not
yield a direct access to physically relevant quantities as the quasienergy or resonance frequencies.
In this paper we have tried to make a first step towards the physical analysis of the CHE solution.

We have addressed essentially two questions that belong to this analysis, the complete time evolution and the explicit analytical form of the quasienergy.
For this purpose we have exploited the fact
that the RPL Schr\"odinger equation has certain symmetries due to the harmonic time dependence of the Hamiltonian.
Accordingly, it is possibe to reduce the time evolution to the first quarter-period.

What remains to be done? Recall that the Floquet theory yields a factorization of the time evolution into a periodic
and an exponential part, the latter involving the quasienergy. In this paper we have only investigated the second part;
but the periodic part including its Fourier coefficients should be also expressible in terms of the series coefficients
of the corresponding CHE solutions. Also the issue of resonance frequencies has not yet been treated in this paper.

Another, mainly mathematical problem would be to apply the techniques used in this paper to  solve the connection problem
for our special CHE in a more explicit way compared with \cite{SS80} and possibly also for a larger class of differential equations.
W.~r.~t.~the needs of physics it would also be desirable to analyze the connection between the CHE solution and the various limit cases of the RPL
known from the literature. We have obtained a first result in this direction by deriving the known limit (\ref{L1}) of the quasienergy
directly from the CHE solution in section \ref{sec:L}.

\section*{Acknowledgment}
This work was funded by the Deutsche Forschungsgemeinschaft (DFG) Grants No.~SCHN 615/25-1 and No.~HO 1771/8-1.
We sincerely thank the members of the DFG Research Unit FOR2692 for fruitful discussions.


\begin{thebibliography}{99}

\bibitem{RabiEtAl54}
	I. Rabi, N. F. Ramsey, and J. Schwinger,
	Use of rotating coordinates in magnetic resonance problems,
	Rev. Mod. Phys. {\bf 26}, 167 (1954).	
	
\bibitem{BlochSiegert40}
	F. Bloch and A. Siegert,
	Magnetic resonance for nonrotating fields,
	Phys. Rev. {\bf 57}, 522 (1940).
	
\bibitem{AutlerTownes55}
	S. Autler and C. H. Townes,
	Stark effect in rapidly varying fields,
	Phys. Rev. {\bf 100}, 703 (1955).
	
\bibitem{Mollow69}
	B. R. Mollow,
	Power spectrum of light scattered by two-level systems,
	Phys. Rev. {\bf 188}, 1969 (1969).
	
\bibitem{GroveEtAl77}
	R. E. Grove, F. Y. Wu, and S. Ezekiel,
	Measurement of the spectrum of resonance fluorescence from a
	two-level atom in an intense monochromatic field,
	Phys. Rev. A {\bf 15}, 227 (1977).	
	
\bibitem{Shirley65}
	J. H. Shirley,
	Solution of the Schr\"odinger equation with a Hamiltonian periodic in
	time,
	Phys. Rev. {\bf 138} (1965).

\bibitem{Floquet83}
	G. Floquet,
	Sur les \'equations diff\'erentielles lin\'eaires \`a coefficients
	p\'eriodiques,
	Annales de l' \'Ecole Normale Sup\'erieure {\bf 12}, 47 (1883).

\bibitem{YakubovichStarzhinskii75}
	V. A. Yakubovich and V. M. Starzhinskii,		
	{\em Linear differential equations with periodic coefficients\/},
	2~volumes (Wiley, New York, 1975).

\bibitem{Zeldovich66}
	Ya. B. Zel'dovich,
	The quasienergy of a quantum-mech\-anical system subjected to a
	periodic action,
	J. Exptl. Theoret. Phys. (U.S.S.R.) {\bf 51}, 1492 (1966)
	[Sov. Phys. JETP {\bf 24}, 1006 (1967)].
	
\bibitem{Ritus66}
	V. I. Ritus,
	Shift and splitting of atomic energy levels by the field of an
	electromagnetic wave,
	J. Exptl. Theoret. Phys. (U.S.S.R.) {\bf 51}, 1544 (1966)
	[Sov. Phys. JETP {\bf 24}, 1041 (1967)].

\bibitem{Sambe73}
	H. Sambe,	
	Steady states and quasienergies of a quantum-mechanical system
	in an oscillating field,
	Phys. Rev. A {\bf 7}, 2203 (1973).
	
\bibitem{FainshteinEtAl78}
	A. G. Fainshtein, N. L. Manakov, and L. P. Rapoport,
	Some general properties of quasi-energetic spectra of quantum systems
	in classical monochromatic fields,
	J. Phys. B: Atom. Molec. Phys. {\bf 11}, 2561 (1978).
	
\bibitem{HillEtAl16}
M.~Shiddiq, D.~Komijani, Y.~Duan, A.~Gaita-Ari\~{n}o, E.~Coronado, and S.~Hill,
Enhancing coherence in molecular spin qubits via atomic clock transitions,
Nature {\bf 531}, 348 -- 351, (2016)


\bibitem{AllenEberly75}
	L. Allen and J. H. Eberly,
	{\em Optical resonance and two-level atoms\/}
	(John Wiley \& Sons, Inc., New York, 1975).


\bibitem{ML07}
T.~Ma, S.-M.~Li,
Floquet system, Bloch oscillation, and Stark ladder,
arXiv:0711.1458v2 [cond-mat.other] (2007)

	
\bibitem{XieHai10}
	Q. Xie and W. Hai,
	Analytical results for a monochromatically driven two-level system,
	Phys. Rev. A {\bf 82}, 032117 (2010).
	




\bibitem{R80}
M.~Razavy,
An exactly soluble Schr\"odinger equation with a bistable potential
\textit{Am. J. Phys.} {\bf 48}, 285 --288 (1980)


\bibitem{JR10}
P.~K.~Jha and Y.~V.~Rostovtsev,
Coherent excitation of a two-level atom driven by a far-off-resonant classical field: Analytical solutions,
\textit{Phys. Rev. A} {\bf 81}, 033827 (2010)

\bibitem{JR10a}
P.~K.~Jha and Y.~V.~Rostovtsev,
Analytical solutions for a two-level system driven by a class of chirped pulses,
\textit{Phys. Rev. A} {\bf 82}, 015801 (2010)


\bibitem{IG14}
A.~M.~Ishkhanyan and A.~E.~Grigoryan,
Fifteen classes of solutions of the quantum two-state problem
in terms of the confluent Heun function,
\textit{Phys. Rev. A} {\bf 47}, 465205 (2014)


\bibitem{ISI15}
A.~M.~Ishkhanyan, T.~A.~Shahverdyan, and T.~A.~Ishkhanyan,
Thirty five classes of solutions of the quantum time-dependent
two-state problem in terms of the general Heun functions,
\textit{Eur. Phys. J. D} {\bf 69}, 10 (2015)


\bibitem{ZXBL13}
H.~Zhong, Q.~Xie, A.~T.~Batchelor, and C.~Lee,
Analytical eigenstates for the quantum Rabi model,
\textit{ J. Phys. A: Math. Theor.} {\bf 46}, 415302 (2013)

\bibitem{MPS14}
A.~J.~Maciejewski, M.~Przybylska, and T.~Stachowiakc,
Full spectrum of the Rabi model,
\textit{Phys. Lett. A} {\bf 378}, 16 -- 20 (2014)


\bibitem{XZBL17}
Q.~Xie, H.~Zhong, A.~T.~Batchelor, and C.~Lee,
The quantum Rabi model: solution and dynamics,
\textit{ J. Phys. A: Math. Theor.} {\bf 50}, 113001 (2017)


\bibitem{X18}
Q.~Xie,
Analytical results for periodically-driven two-level models
in relation to Heun functions,
\textit{Pramana -- J. Phys.} {\bf 91}, 19 (2018)



\bibitem{S18}
H.-J.~Schmidt,
The Floquet theory of the two level system revisited,
\textit{Z. Naturforsch. A} {\bf 73} (8), 705 -- 731 (2018)


\bibitem{H16}
M.~Holthaus,
Floquet engineering with quasienergy bands of periodically driven optical lattices,
\textit{J. Phys. B: At. Mol. Opt. Phys.} {\bf  49}, 013001 (2016)






\bibitem{DLMF}   
NIST Digital Library of Mathematical Functions. http://dlmf.nist.gov/, Release 1.0.19 of 2018-06-22. F. W. J. Olver, A. B. Olde Daalhuis, D. W. Lozier, B. I. Schneider, R. F. Boisvert, C. W. Clark, B. R. Miller, and B. V. Saunders, eds.



\bibitem{SS80}
R.~Sch\"afke and D.~Schmidt,
The connection problem for general linear ordinary differential equations at two regular singuar points with applications in the theory of special functions,
\textit{SIAM J. Math. Anal.} {\bf 11} (5), 848 -- 862, (1980)

\bibitem{S06}
S.~Schultze,
Zur globalen Theorie der konfluenten Heun\-schen Differentialgleichung,
Dissertation, Universit\"at Duisburg-Essen (2006)



\bibitem{Holthaus92}
	M. Holthaus,
	The quantum theory of an ideal superlattice responding to
	far-infrared laser radiation,
	Z. Phys. B {\bf 89}, 251 (1992).
	
\bibitem{ZenesiniEtAl09}
	A. Zenesini, H. Lignier, D. Ciampini, O. Morsch, and E. Arimondo,
	Coherent control of dressed matter waves,
	Phys. Rev. Lett. {\bf 102}, 100403 (2009).
	
\bibitem{StruckEtAl11}
	J. Struck, C. \"Olschl\"ager, R. Le Targat, P. Soltan-Panahi,
	A. Eckardt, M. Lewenstein, P. Windpassinger, and K. Sengstock,
	Quantum simulation of frustrated classical magnetism in triangular
	optical lattices,
	Science {\bf 333}, 996 (2011).
	
\bibitem{AidelsburgerEtAl13}
	M. Aidelsburger, M. Atala, M. Lohse, J. T. Barreiro, B. Paredes,
	and I. Bloch,
	Realization of the Hofstadter Hamiltonian with ultracold
	atoms in optical lattices,
	Phys. Rev. Lett. {\bf 111}, 185301 (2013).
	
\bibitem{MiyakeEtAl13}
	H. Miyake, G. A. Siviloglou, C. J. Kennedy, W. C. Burton,
	and W. Ketterle,
	Realizing the Harper Hamiltonian with laser-assisted tunneling
	in optical lattices,
	Phys. Rev. Lett. {\bf 111}, 185302 (2013).
				
\bibitem{Eckardt17}
	A. Eckardt,
	Colloquium: Atomic quantum gases in periodically driven optical
	lattices,
	Rev. Mod. Phys. {\bf 89}, 011004 (2017).
	
\bibitem{GorgEtAl18}
	F. G\"org, M. Messer, K. Sandholzer, G. Jotzu, R. Desbuquois,
	and T. Esslinger,
	Enhancement and sign change of magnetic correlations in a driven
	quantum many-body system,
	Nature {\bf 553}, 481 (2018).


\end{thebibliography}
\end{document}